\documentclass[prl,preprint,groupedaddress, showkeys,showpacs]{revtex4}

\usepackage{graphicx}
\begin{document}


\title{Thermal noise can facilitate energy transformation in the presence of entropic barriers }


\author{Bao-quan  Ai$^{1}$}\email[Email: ]{aibq@scnu.edu.cn}
\author{Hui-zhang Xie$^{2}$}
\author{Liang-gang Liu$^{3}$}

\affiliation{$^{1}$ Institute for Condensed Matter Physics, School
of Physics and Telecommunication Engineering and Laboratory of
Photonic Information Technology, South China
Normal University, 510006 Guangzhou, China\\
$^{2}$Department of Physics, South China University of technology, 510641 Guangzhou, China\\
$^{3}$ Faculty of Information Technology , Macau University of
Science and Technology, Macao}


\date{\today}
\begin{abstract}
\indent Efficiency of a Brownian particle moving along the axis of a
three-dimensional asymmetric periodic channel is investigated in the
presence of a symmetric unbiased force and a load. Reduction of the
spatial dimensionality from two or three physical dimensions to an
effective one-dimensional system entails the appearance of entropic
barriers and an effective diffusion coefficient. The energetics in
the presence of entropic barriers exhibits peculiar behavior which
is different from that occurring through energy barriers. We found
that even on the quasistatic limit there is a regime where the
efficiency can be a peaked function of temperature, which indicates
that thermal noise can facilitate energy transformation, contrary to
the case of energy barriers. The appearance of entropic barriers may
induce optimized efficiency at a finite temperature.

\end{abstract}

\pacs{05. 06. Cd, 02. 50. Ey, 05. 40. Jc, 66. 10. Cb }



\maketitle


\section {Introduction}

\indent Noise-induced transport play a crucial role in many
processes from physical and biological to social systems. There has
been an increasing interest in transport properties of nonlinear
systems which can extract usable work from unbiased nonequilibrium
fluctuations \cite{1,2,3,4}. This comes from the desire of
understanding molecular motors \cite{5}, nanoscale friction
\cite{6}, surface smoothening \cite{7}, coupled Josephson junctions
\cite{8}, optical ratchets and directed motion of laser cooled atoms
\cite{9}, and mass separation and trapping schemes at microscale
\cite{10}.

\indent Most studies have referred to the consideration of the
energy barriers. The nature of the energy barriers depends on which
thermodynamic potential varies when passing from one well to the
other, and its presence plays an important role in the dynamics of
the solid-state physics system. However, in some cases, such as soft
condensed-matter and biological systems, the entropic barriers
should be considered. The entropic barriers may appear when
coarsening the description of a complex system  for simplifying its
dynamics. Reguera and co-workers \cite{11,12} used the mesoscopic
nonequilibrium thermodynamics theory to derive the general kinetic
equation of a system and analyze in detail the case of diffusion in
a domain of irregular geometry in
 which the presence of the boundaries induces entropic barriers
 when approaching the dynamics by a coarsening of the description.
 In the presence of entropic barriers, the asymmetry of the tube can induce a net current in
 the absence of any net macroscopic forces or in the presence of
 the unbiased forces \cite{13}. \\
\indent In recent years the energetics of these systems, which
rectify the zero-mean fluctuations, was investigated.  Much of the
interest was motivated by the elegant piece of work done by Magnasco
\cite{14}, which showed that a Brownian particle, subject to
external fluctuations, can undergo a non-zero drift while moving
under the influence of an asymmetric potential. He claimed that
there is a region where the efficiency can be optimized at a finite
temperature and the existence of thermal fluctuations facilitate the
efficiency of energy transformation. His claim is interesting
because thermal noise is usually known to disturb the operation of
machines. Based on energetic analysis of the same model Kamegawa and
co-workers \cite{15} made a important conclusion that the efficiency
of energy transformation on the quasistatic limit cannot be
optimized at finite temperatures and that the thermal fluctuations
does not facilitate it. Takagi and Hondou\cite{16} found that
thermal noise may facilitate the energy conversion in the forced
thermal ratchet when an "oscillating ratchet" was considered. A
recent investigation of Dan and co-workers \cite{17} showed that the
efficiency can be optimized at a finite temperature in inhomogeneous
systems with spatially varying friction coefficient in a forced
thermal ratchet. Sumithra and co-workers \cite{18} studied a
homogeneous ratchet driven by a nonadiabatical external periodic
force and found that thermal noise can facilitate the energy
transformation. When an isothermal ratchet driven by a chemical
reaction between the two states, the efficiency can also have a
maximum as a function of temperature \cite{19}. In forced undamped
ratchets \cite{20}, the efficiency optimization can also be found.
In two-dimensional ratchet, Wang and Bao \cite{21} found that the
efficiency is peaked function of temperature which is different from
that in one-dimensional ratchet. Recently, Ghosh and co-workers
\cite{22} investigated the stochastic energetics of direction
quantum transport due to rectification of nonequilibrium thermal
fluctuations and found that the Stokes efficiency reaches a maximum
at a particular temperature. Efficiency optimization in forced
thermal ratchet usually takes place for fast oscillating or
fluctuation forces. When the force changes very slowly enough, for
example a square wave with a very long period, the system can be
treated as quasistatic. When the unbiased force is asymmetry, Ai and
co-workers \cite{4} found that the thermal noise can facilitate the
energy transformation even on the quasi-static limit.

\indent The previous works on efficiency were on the consideration
of the energy barriers. The present work is extended to the study of
efficiency to the case of entropic barriers. We emphasize on finding
whether the thermal noise in the presence of entropic barriers can
facilitate the energy transformation even on the quasistatic
limit.\\

\section {Efficiency in a three-dimensional periodic tube}

\begin{figure}[htbp]
  \begin{center}\includegraphics[width=10cm,height=6cm]{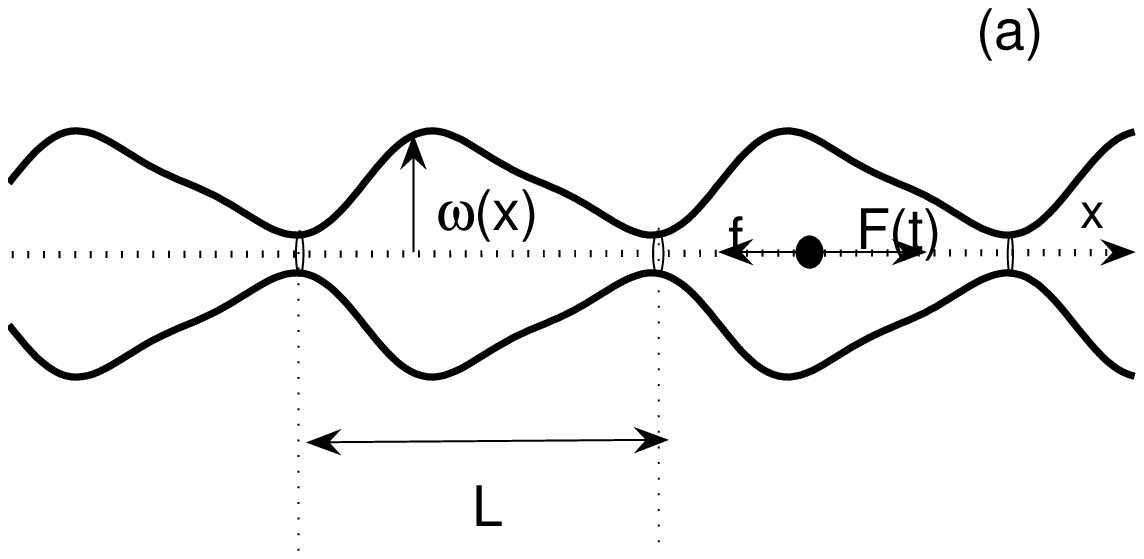}
 \includegraphics[width=10cm,height=6cm]{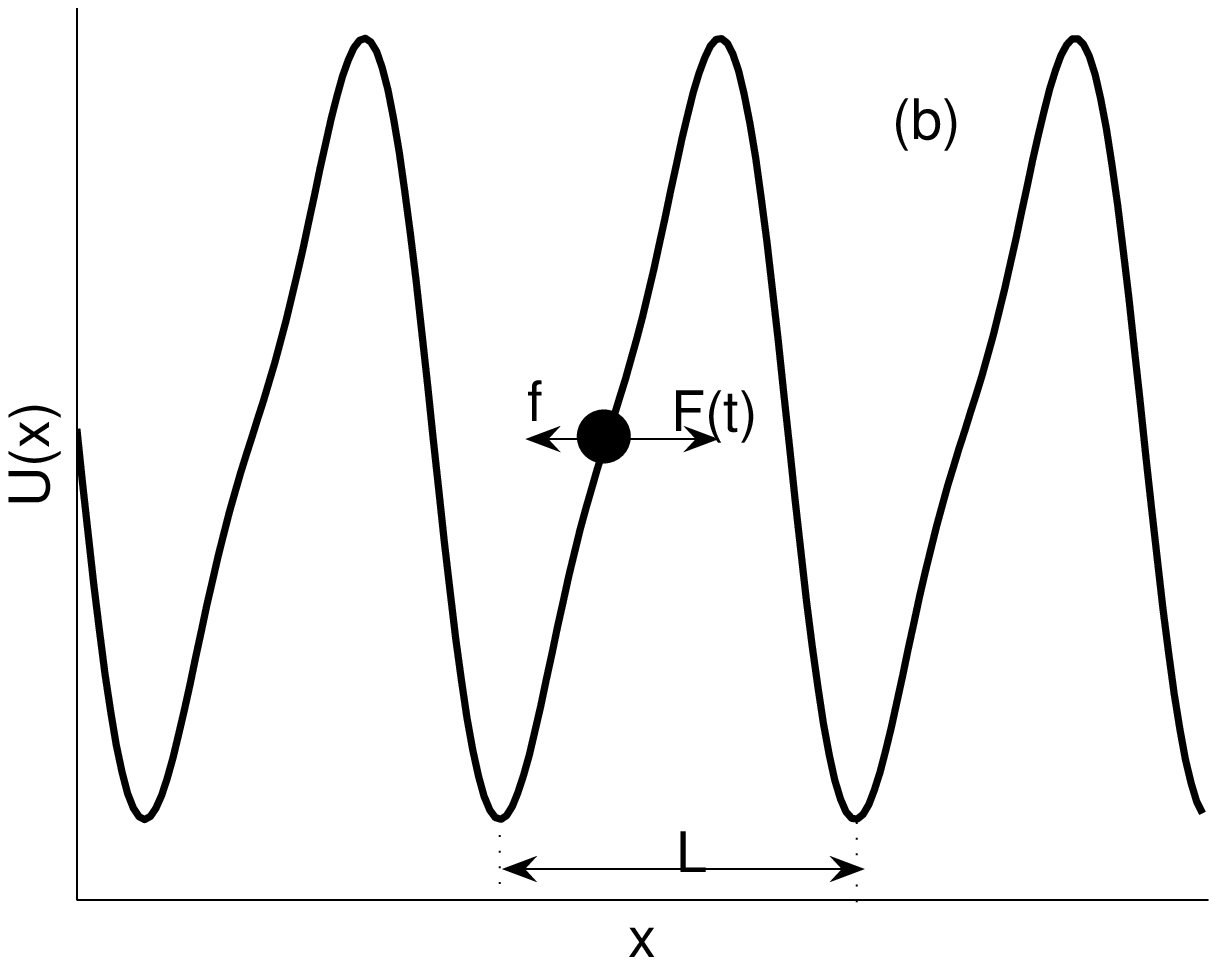}
  \caption{ \baselineskip 0.4in Schematic diagrams of a tube and a conventional energetic ratchet. (a) A tube with periodicity $L$. The half-width $\omega(x)$ is a periodic function of $x$, $ \omega(x)=a[\sin(\frac{2\pi x}{L})+\frac{\Delta}{4}\sin(\frac{4\pi
    x}{L})]+b$. $\Delta$ is the asymmetric parameter of the tube shape. $F(t)$ and $f$ are the zero-mean periodic force and the external load, respectively.
    (b)The conventional energetic ratchet with periodicity $L$. $U(x)=-U_{0}[\sin(\frac{2\pi x}{L})+\frac{\Delta}{4}\sin(\frac{4\pi x}{L})]$.
    $U_{0}$ is amplitude of the potential and $\Delta$ the asymmetric parameter of the potential. $F(t)$ and $f$ are the same as that in (a).}\label{1}
\end{center}
\end{figure}

\indent We consider an overdamped Brownian particle moving in an
asymmetric periodic tube [Fig. 1(a)] in the presence of a symmetric
unbiased external force and a load. Its overdamped dynamics is
described by the following Langevin equations\cite{11,12,13}

\begin{equation}\label{}
    \gamma\frac{dx}{dt}=F(t)-f+\xi_{x}(t),
\end{equation}
\begin{equation}\label{}
    \gamma\frac{dy}{dt}=\xi_{y}(t),
\end{equation}
\begin{equation}\label{}
    \gamma\frac{dz}{dt}=\xi_{z}(t),
\end{equation}
where $x$, $y$, $z$  are the three-dimensional (3D) coordinates,
$\gamma$ is the friction coefficient of the particle.
$\xi_{x,y,z}(t)$ are the uncorrelated Gaussian white noises with
zero mean and correlation function:
$<\xi_{i}(t)\xi_{j}(t^{'})>=2\gamma
k_{B}T\delta_{i,j}\delta(t-t^{'})$ for $i,j=x, y, z$. $k_{B}$ is the
Boltzmann constant and  $T$ is the absolute temperature. $<...>$
denotes an ensemble average over the distribution of noise.
$\delta(t)$ is the Dirac delta function. Imposing reflecting
boundary conditions in the transverse direction ensures the
confinement of the dynamics within the tube, while periodic boundary
conditions are enforced along the longitudinal direction. $f$ is a
load against the Brownian motor along the $x$ direction. The shape
of the tube is described by its radius,
\begin{equation}\label{}
    \omega(x)=a[\sin(\frac{2\pi x}{L})+\frac{\Delta}{4}\sin(\frac{4\pi
    x}{L})]+b,
\end{equation}
where $a$ is the parameter that controls the slope of the tube,
$\Delta$ is the asymmetry parameter of the tube shape and $L$ is its
periodicity. The radius at the bottleneck is
$r_{b}=b-a(1+\frac{\Delta}{4})$. $F(t)$ is a temporally symmetric
unbiased external force and satisfies
\begin{equation}\label{}
 F(t)=\left\{
\begin{array}{ll}
   F_{0},& \hbox{$n\tau\leq
t<n\tau+\frac{1}{2}\tau$};\\
   -F_{0} ,&\hbox{$n\tau+\frac{1}{2}\tau<t\leq
      (n+1)\tau$},\\
\end{array}
\right.
\end{equation}
where $\tau$ is the period of the unbiased force and $F_{0}$ is its
magnitude.

\indent The movement equation of a Brownian particle moving along
the axis of the 3D (or 2D) tube can be described by the Fick-Jacobs
equation \cite{11,12,13,23,24} which is  derived from the 3D (or 2D)
Smoluchowski equation after elimination of $y$ and $z$ coordinates
by assuming equilibrium in the orthogonal directions. The
complicated boundary conditions of the diffusion equation in
irregular channels can be greatly simplified by introducing an
entropic potential that accounts for the reduced space accessible
for diffusion of the Brownian particle. Reduction of the coordinates
may involve the appearance of entropic barriers and an effective
diffusion coefficient. When $|\omega^{'}(x)|<<1$, the effective
diffusion coefficient reads \cite{11,12,13,23,24}
\begin{equation}\label{}
    D(x)=\frac{D_{0}}{[1+\omega^{'}(x)^{2}]^{\alpha}},
\end{equation}
where $D_{0}=k_{B}T/\gamma$ and $\alpha=1/3$, $1/2$ for two and
three dimensions, respectively. The prime stands for the derivative
with respect to the space variable $x$.

\indent Consider the effective diffusion coefficient and the
entropic barriers, the dynamics of a Brownian particle moving along
the axis of the 3D (or 2D) tube can be described by the equation
\cite{11,12,13}
\begin{equation}\label{}
    \frac{\partial P(x,t)}{\partial t}=\frac{\partial}{\partial x}[D(x)\frac{\partial P(x,t)}{\partial
    x}+\frac{D(x)}{k_{B}T}\frac{\partial A(x,t)}{\partial x}P(x,t)]=-\frac{\partial j(x,t)}{\partial
    x},
\end{equation}
where a free energy $A(x,t)=E-TS=fx-F(t)x-Tk_{B}\ln h(x)$ is defined
\cite{11}: here $E=fx-F(t)x$ is the energy, $S=k_{B}\ln h(x)$ is the
entropy, $h(x)$ is the dimensionless width $2\omega(x)/L$ in two
dimensions, and the dimensionless transverse cross section
$\pi[\omega(x)/L]^{2}$ of the tube in three dimensions. $j(x,t)$ is
the probability current density. $P(x,t)$ is the probability density
for the particle at position $x$ and at time $t$. It satisfies the
normalization condition and the periodicity condition
\begin{equation}\label{}
    \int_{0}^{L}P(x,t)dx = 1, \indent
P(x,t)= P(x+L,t).
\end{equation}
\indent If $F(t)$ changes very slowly with respect to $t$, namely,
its period is longer than any other time scale of the system,
 there exists a quasistatic state.  It is noted that the validity of the Fick-Jacobs
 equation in the limit $T\rightarrow 0$ is questionable \cite{24}
 and the quasistatic approximate method would also fail. However,
 our study is focus on the intermediate values of temperature, and
 thus our method is still valid.  In this case, by following the  method in \cite{1,2,3,4,5,6,11,12,13}, we can obtain the
current
\begin{equation}\label{}
    j(F(t))=\frac{k_{B}T[1-\exp(-\frac{(F(t-f))L}{k_{B}T})]}{\int_{0}^{L}h(x)\exp(\frac{(F(t)-f)x}{k_{B}T})dx\int_{x}^{x+L}[1+\omega^{'}(y)^{2}]^{\alpha}h^{-1}(y)\exp(-\frac{(F(t)-f)y}{k_{B}T})dy}.
\end{equation}

For the given external force $F(t)$ (Eq. (5)), the average current
over a period is obtained,
\begin{equation}\label{9}
    J=\frac{1}{\tau}\int_{0}^{\tau}j(F(t))dt=\frac{1}{2}[j(F_{0})+j(-F_{0})],
\end{equation}
where $j(F_{0})$ is a current induced by a constant force
$F(t)=F_{0}$.

 \indent According to the energetics \cite{25}, the input energy $E_{in}$ per unit time from the external
force to the system is given by \cite{15,25},
\begin{equation}\label{9}
E_{in}=\frac{1}{t_{j}-t_{i}}\int^{x(t_{j})}_{x(t_{i})}F(t)dx(t).
\end{equation}
From Eq. (9), the input energy per unit time is
\begin{equation}\label{11}
  E_{in}=\frac{1}{2}F_{0}[j(F_{0})-j(-F_{0})].
\end{equation}
\indent In order for the system to do useful work, the load force
$f$ is applied in a direction opposite to the direction of current
in the system. If the current flows in the same direction as the
load, no useful work is done. The average work per unit time done
over a period is given by
\begin{equation}\label{12}
W=\frac{1}{2}f[j(F_{0})+j(-F_{0})].
\end{equation}
\indent Thus, the efficiency $\eta$ of the system to transform the
external fluctuation to useful work is given by \cite{15,25}
\begin{equation}\label{13}
  \eta=\frac{W}{E_{in}}=\frac{f[j(F_{0})+j(-F_{0})]}{F_{0}[j(F_{0})-j(-F_{0})]}.
\end{equation}

\indent Because of $\frac{j(-F_{0})}{j(F_{0})}<0$, Eq. (14) can be
rewritten
\begin{equation}\label{}
    \eta=\frac{f}{F_{0}}[1-\frac{2|\frac{j(-F_{0})}{j(F_{0})}|}{1+|\frac{j(-F_{0})}{j(F_{0})}|}].
\end{equation}
\indent  Equation (15) shows that the efficiency $\eta$ depends on
$|\frac{j(-F_{0})}{j(F_{0})}|$.  When
$|\frac{j(-F_{0})}{j(F_{0})}|\rightarrow 0$, the maximum
efficiency of the energy transformation is $\eta_{max}=f/F_{0}$.

\section {Results and discussions}
\indent  Because the results  from two and three dimensions are very
similar, for the convenience of physical discussion, we now mainly
investigated the energetics in three dimensions with $k_{B}=1$ and
$\gamma=1$. In order to compare with the conventional energetic
ratchet \cite{15}, we also investigate the current and efficiency in
the presence of asymmetric energy barriers on the quasistatic
limit[Fig. 1 (b)].

\begin{figure}[htbp]
  \begin{center}\includegraphics[width=10cm,height=8cm]{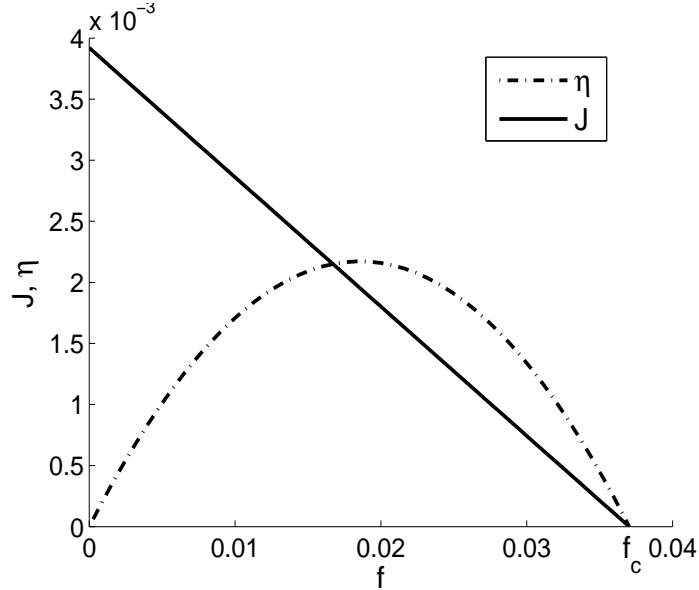}
  \caption{Efficiency $\eta$ and current $J$ vs the load $f$ at $a=\frac{1}{2\pi}$, $b=\frac{1.5}{2\pi}$, $L=2\pi$, $\alpha=1/2$, $F_{0}=0.5$, $\Delta=1.0$, and $T=0.5$.
  The solid and dash-dotted line denotes $J$ and $\eta$, respectively.}\label{1}
\end{center}
\end{figure}
\indent In Fig. 2, we plot the efficiency $\eta$ and current $J$ as
a function of the load $f$ in the present system.  It is expected
that the efficiency exhibits a maximum as a function of the load. It
is obvious that the efficiency is zero for the case of zero loading.
At the critical value of $f_{c}$ the current is zero and hence the
efficiency also tends to zero. Between $f=0$ and $f=f_{c}$ the
efficiency exhibits a maximum. Beyond $f=f_{c}$ the current flows
down the load and therefore the definition of efficiency becomes
invalid. It is also found that the current is a monotonically
decreasing function of the load.

\begin{figure}[htbp]
  \begin{center}\includegraphics[width=10cm,height=8cm]{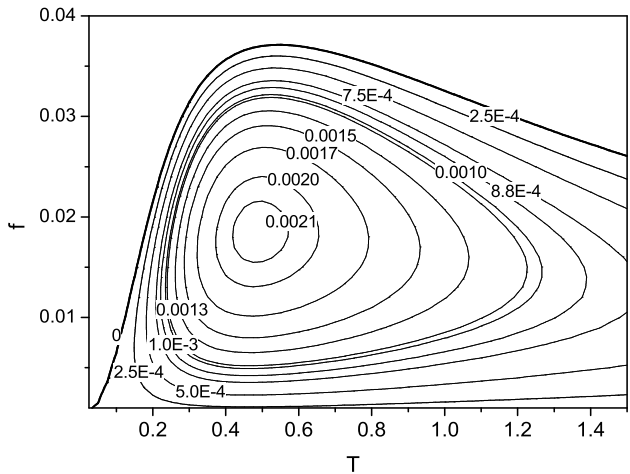}
  \caption{Efficiency contours on $f-T$ plane at $a=\frac{1}{2\pi}$, $b=\frac{1.5}{2\pi}$, $L=2\pi$, $\alpha=1/2$, $F_{0}=0.5$, and $\Delta=1.0$.
  The thick solid line indicates $\eta=0$ contour.}\label{1}
\end{center}
\end{figure}

\indent Figure 3 shows the efficiency contours on $f-T$ plane in the
present system. The thick solid line indicates $\eta=0$ contour.
When the temperature is low, the maximum load (referring to the
critical value of the load $f_{c}$ in Fig. 2) is small. When the
temperature is very high, the maximum load is also small. Therefore
there exists a temperature at which the maximum load takes its
maximum. The efficiency near the center is high and low far from the
center. For a given valid load, the efficiency can be a peaked
function of temperature.

\begin{figure}[htbp]
  \begin{center}\includegraphics[width=7cm,height=6cm]{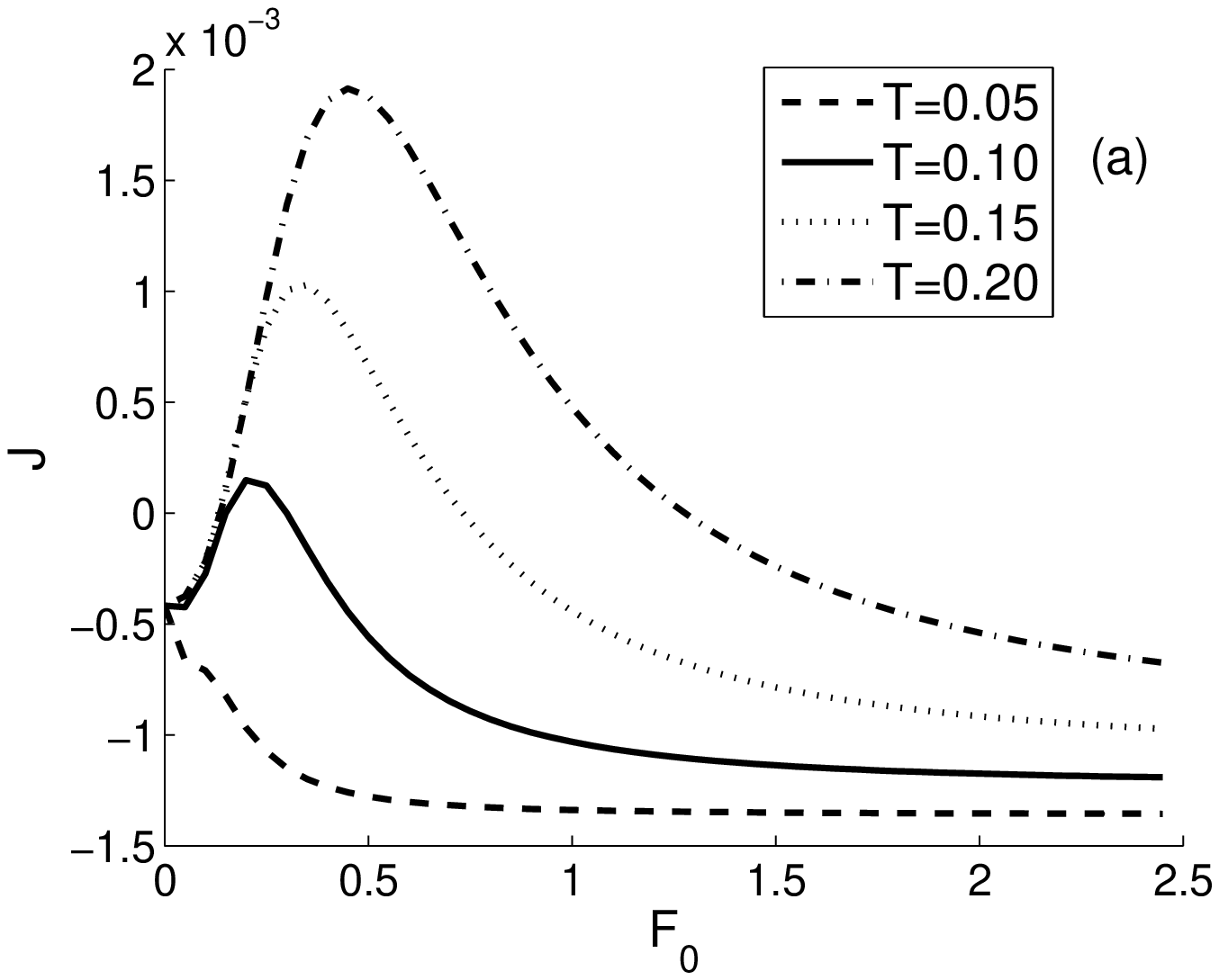}
  \includegraphics[width=7cm,height=6cm]{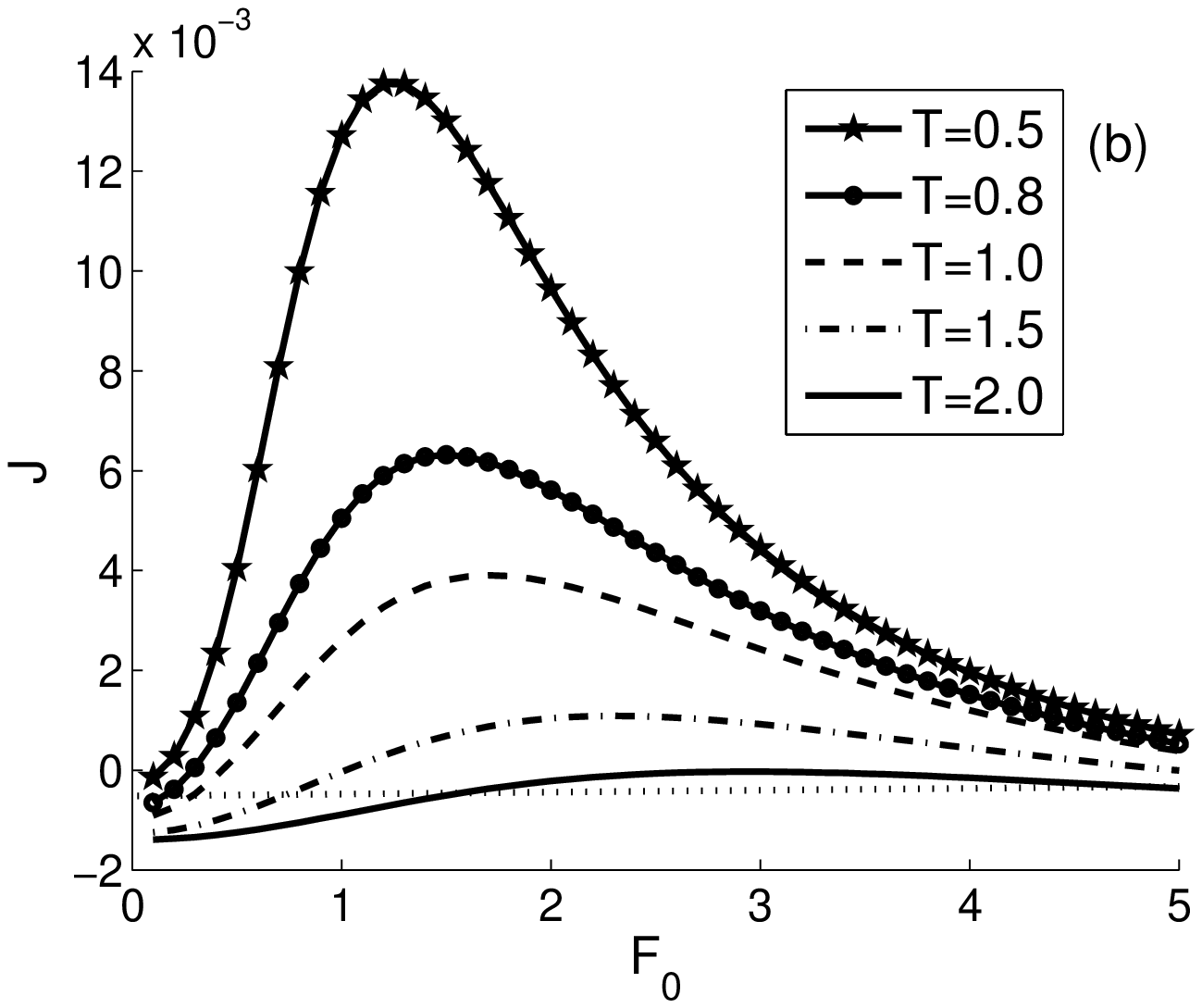}
  \caption{Current $J$ as a function of $F_{0}$ for different values of temperature. (a)Present system  at $a=\frac{1}{2\pi}$, $b=\frac{1.5}{2\pi}$, $L=2\pi$, $\alpha=1/2$, $\Delta=1.0$, and $f=0.01$.
  (b)The conventional energetic ratchet at $U_{0}=2.0$, $\Delta=1.0$, $L=2\pi$, and $f=0.01$. }\label{1}
\end{center}
\end{figure}

\indent Figure 4 shows the current as a function of $F_{0}$ for both
the present system (a) and the conventional ratchet (b).  The
current is a peaked function of $F_{0}$ for both cases.  The current
increases with the temperature in present system [Fig. 4(a)]. The
current tends to negative at low temperatures ($T=0.05$) which
indicates that the current is dominated by the load $f$ for low
temperatures. However, in conventional ratchet the current decreases
with the temperature [Fig. 4(b)]. The current goes to negative for
high temperatures which shows that the load $f$ dominates at high
temperatures ($T=2.0$).

\begin{figure}[htbp]
  \begin{center}\includegraphics[width=7cm,height=6cm]{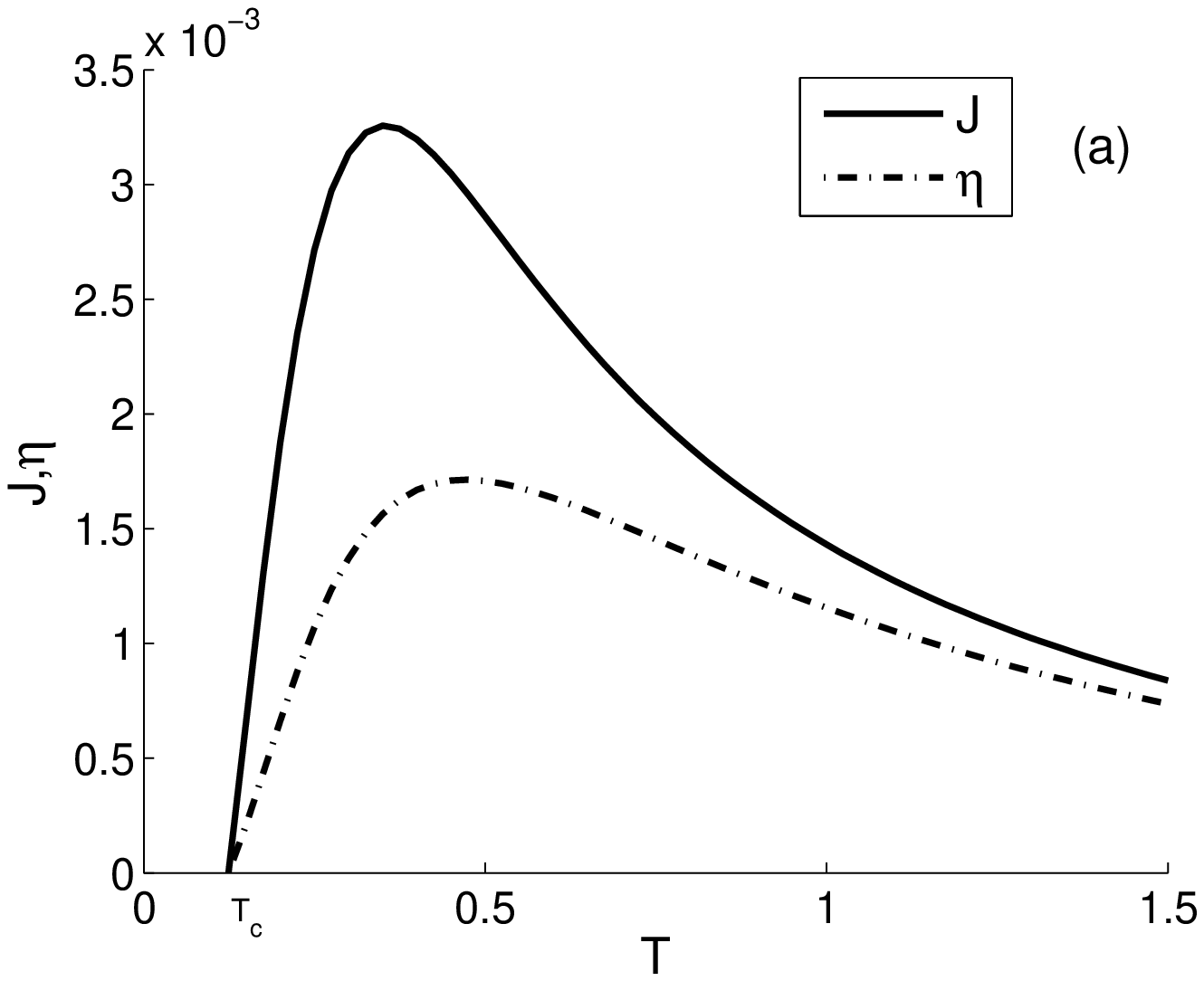}
  \includegraphics[width=7cm,height=6cm]{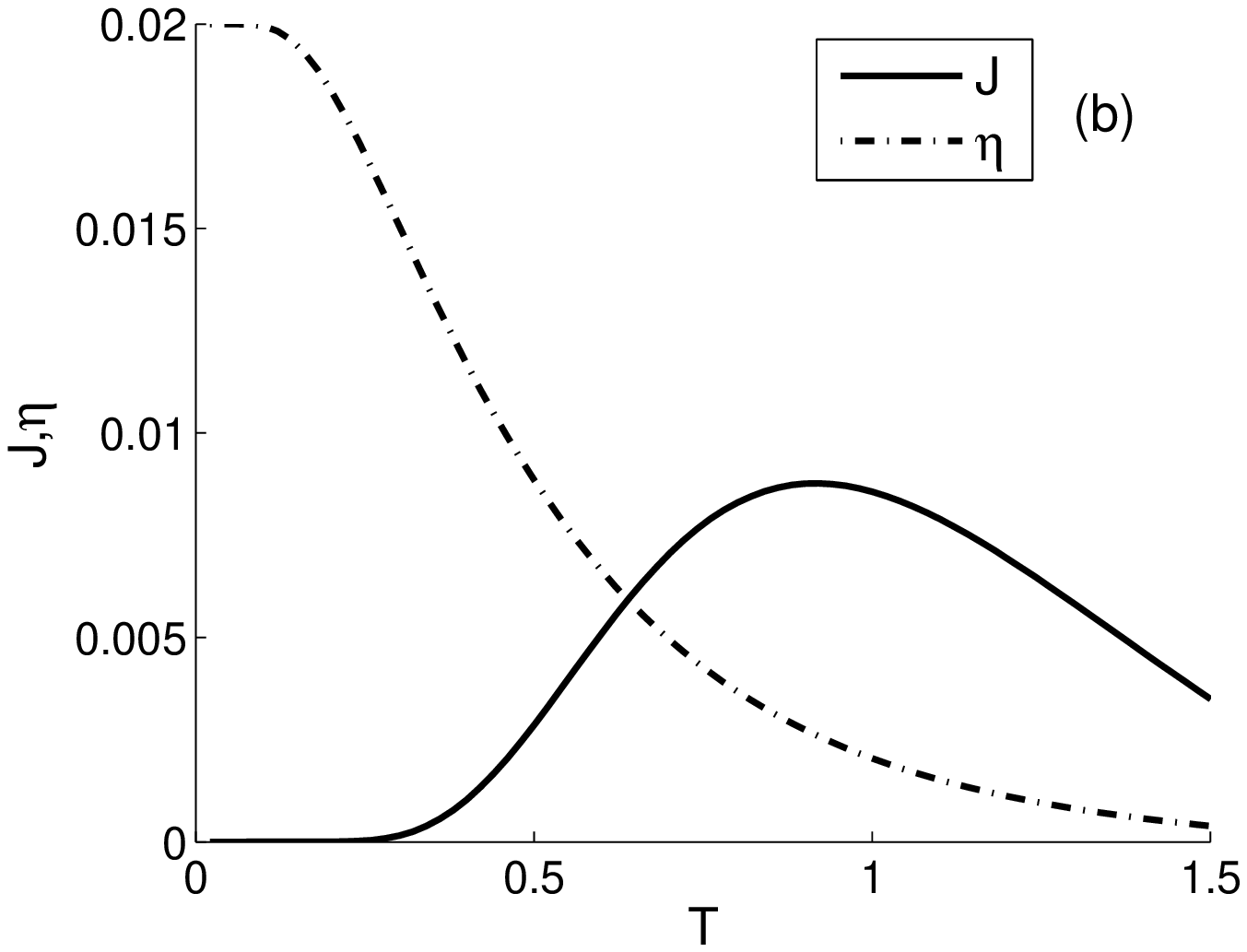}
  \caption{Efficiency $\eta$ and current $J$ as a function of temperature $T$. (a)Present system  at $a=\frac{1}{2\pi}$, $b=\frac{1.5}{2\pi}$, $L=2\pi$, $\alpha=1/2$, $F_{0}=0.5$, $\Delta=1.0$, and $f=0.01$.
  (b)The conventional energetic ratchet at $U_{0}=2.0$, $\Delta=1.0$, $L=2\pi$, $F_{0}=0.5$, and $f=0.01$. }\label{1}
\end{center}
\end{figure}

\indent Figure 5(a) shows the efficiency and current as a function
of temperature in present system. The current and efficiency is
observed to be bell shaped, which shows the feature of resonance.
When $T\rightarrow 0$, the Brownian particle can only move along $x$
axis (1D) and the effect of entropic barriers disappears and the
load dominates. Thus, the current is negative. When $T>T_{c}$, the
entropic barriers take effect, the asymmetry of the tube shape can
induce a net current \cite{13}. When $T\rightarrow \infty$, the
effect of the unbiased force disappears and the load dominates, the
current becomes negative again. There is an optimized value of $T$
at which the current $J$ takes its maximum. The current in
conventional ratchet is similar to that in present system [the solid
line in Fig. 5(a) and 5(b)]. However, in the conventional ratchet
\cite{15} [the dash-dotted line in Fig. 5(b)] the energetic
efficiency of Brownian motors is a monotonically decreasing function
of temperature on the quasistatic limit. In that case, thermal noise
cannot facilitate energy transformation. Contrary to that work, the
efficiency in our system [the dash-dotted line in Fig. 5(a)] is a
peaked function of temperature which indicates that thermal noise
can facilitate energy conversion. For the case of entropic barriers,
even when the current goes to zero at $T=T_{c}$, local current
($j(F_{0})$, $j(-F_{0})$) that refers to the current in each
semiperiod  still remains finite. Thus, there exists finite energy
dissipation at $T=T_{c}$, which shows that the input energy $E_{in}$
still remains finite. Therefore the efficiency is found to zero at
$T=T_{c}$, and takes its maximum at a finite temperature. It is to
be mentioned that the temperature corresponding to the maximum
efficiency is not the same as the temperature at which the current
becomes maximum, which is similar to the case of energetic barriers
\cite{4,17}.

\begin{figure}[htbp]
  \begin{center}\includegraphics[width=7cm,height=6cm]{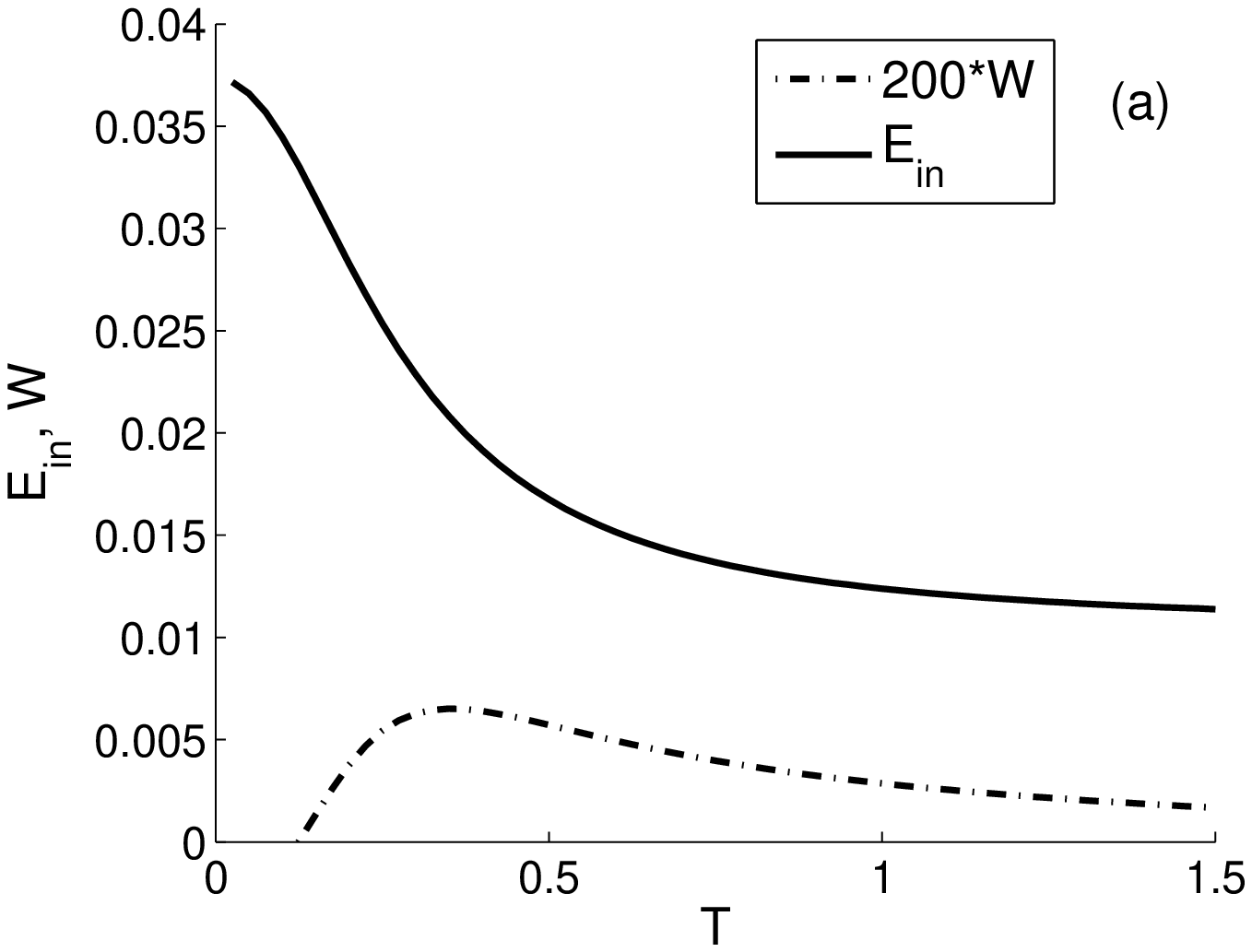}
  \includegraphics[width=7cm,height=6cm]{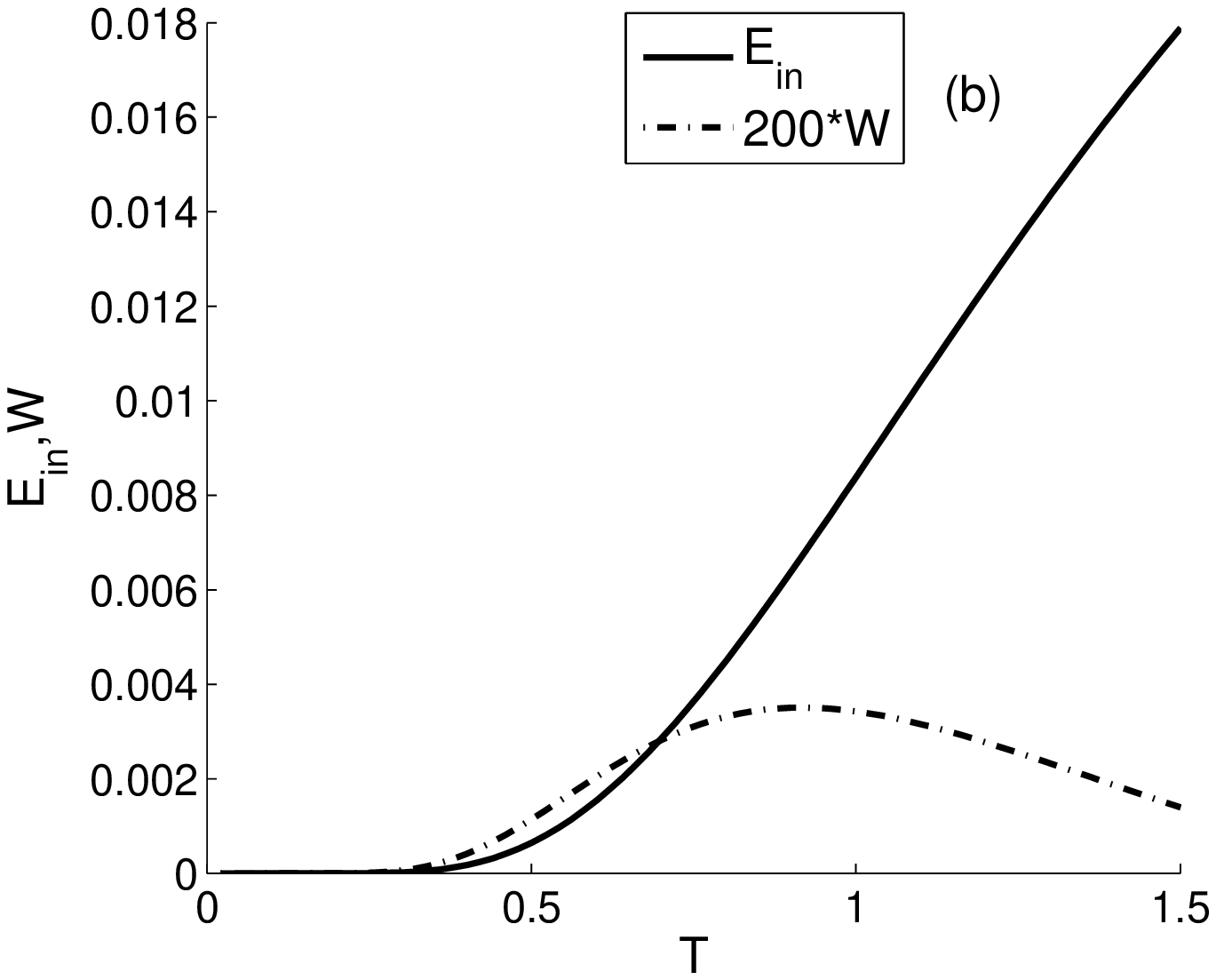}
  \caption{Useful work $W$ and input energy $E_{in}$ vs temperature ($W$ has been scaled up by a factor of 200 to make it comparable with $E_{in}$).  (a)Present system  at $a=\frac{1}{2\pi}$, $b=\frac{1.5}{2\pi}$, $L=2\pi$, $\alpha=1/2$, $F_{0}=0.5$, $\Delta=1.0$, and $f=0.01$.
  (b)The conventional energetic ratchet at $U_{0}=2.0$, $\Delta=1.0$, $L=2\pi$, $F_{0}=0.5$, and $f=0.01$.
  }\label{1}
\end{center}
\end{figure}
\indent  In Fig. 6(a) we plot input energy $E_{in}$ and useful work
$W$ (scaled up by a factor 200 for convenience of comparison) as a
function of temperature. The useful work $W$ has a peak at a finite
temperature, because of the behavior of the current during the
period $\tau$. The input energy $E_{in}$ is a monotonically
decreasing function of temperature. At the limit $T\rightarrow 0$,
$E_{in}$ tends to its maximum, where all input energy dissipates
because in the absence of energetic barriers the unbiased force and
the load make finite local current even at the limit $T\rightarrow
0$. Therefore the efficiency takes its maximum at a finite
temperature. However, in the conventional ratchet [Fig. 6(b)] the
input energy $E_{in}$ is a monotonically increasing function of
temperature. At the limit $T\rightarrow 0$, the current tends to
zero and the input energy also goes to zero, which is different from
that in the present system.

\begin{figure}[htbp]
  \begin{center}\includegraphics[width=7cm,height=6cm]{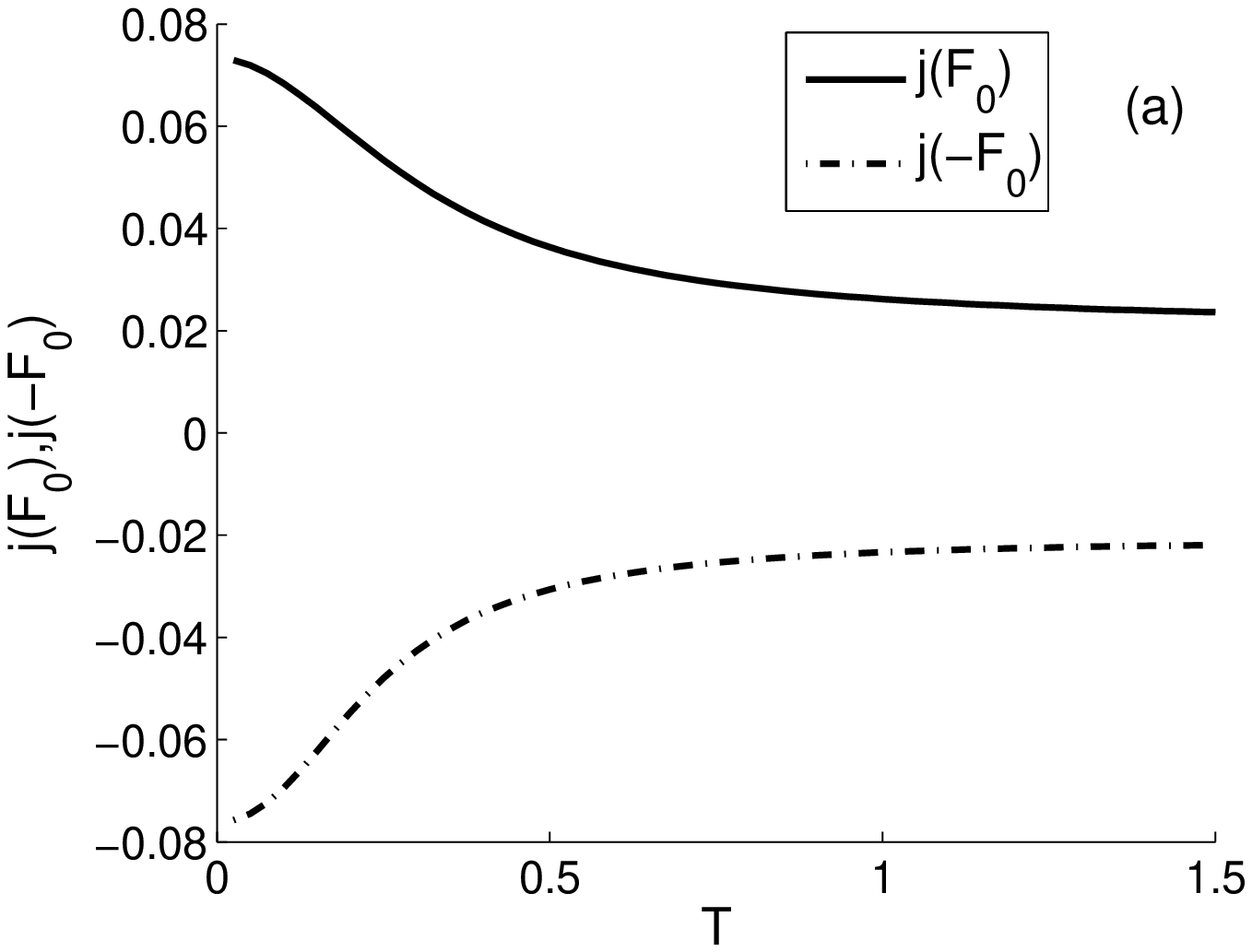}
  \includegraphics[width=7cm,height=6cm]{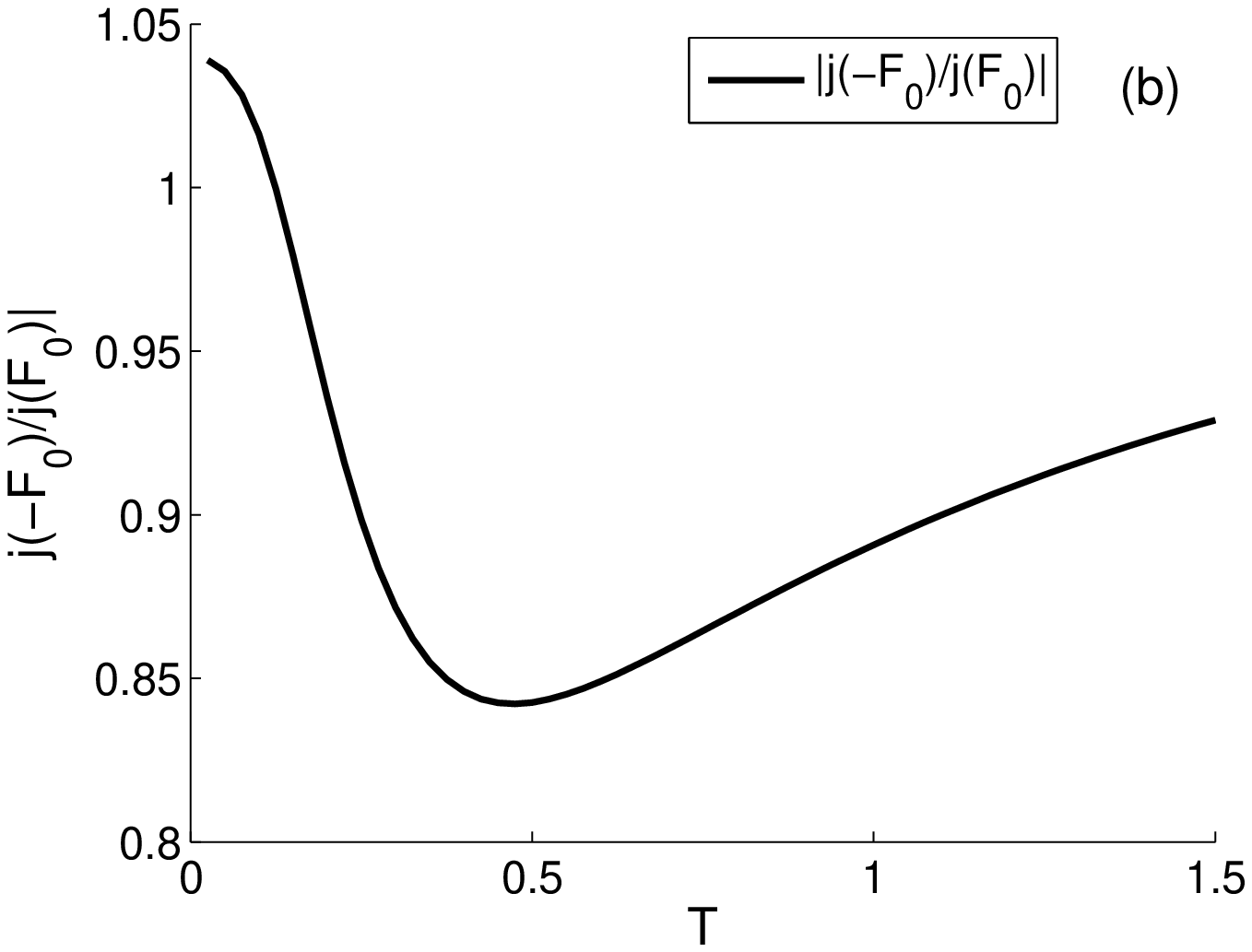}
  \includegraphics[width=7cm,height=6cm]{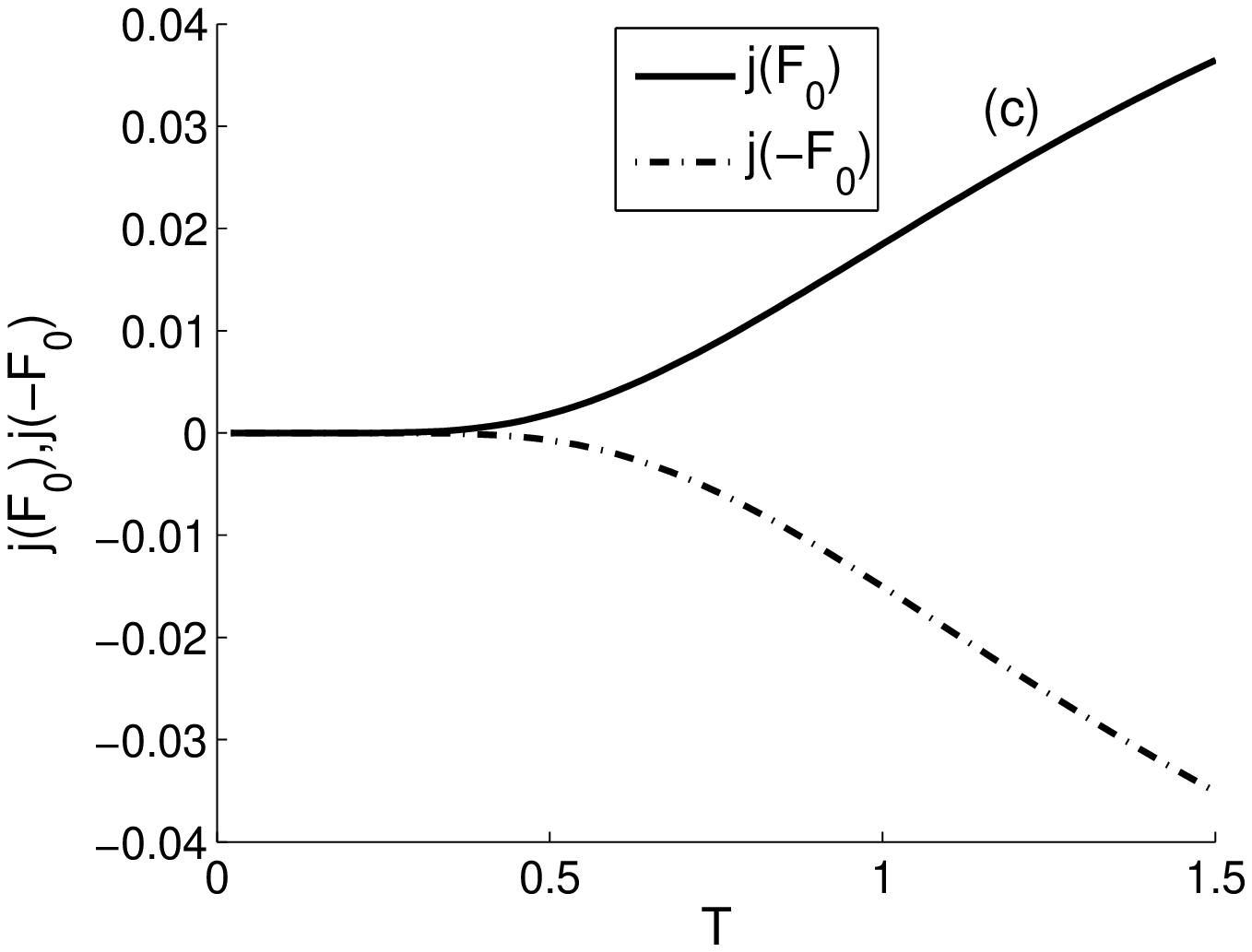}
  \includegraphics[width=7cm,height=6cm]{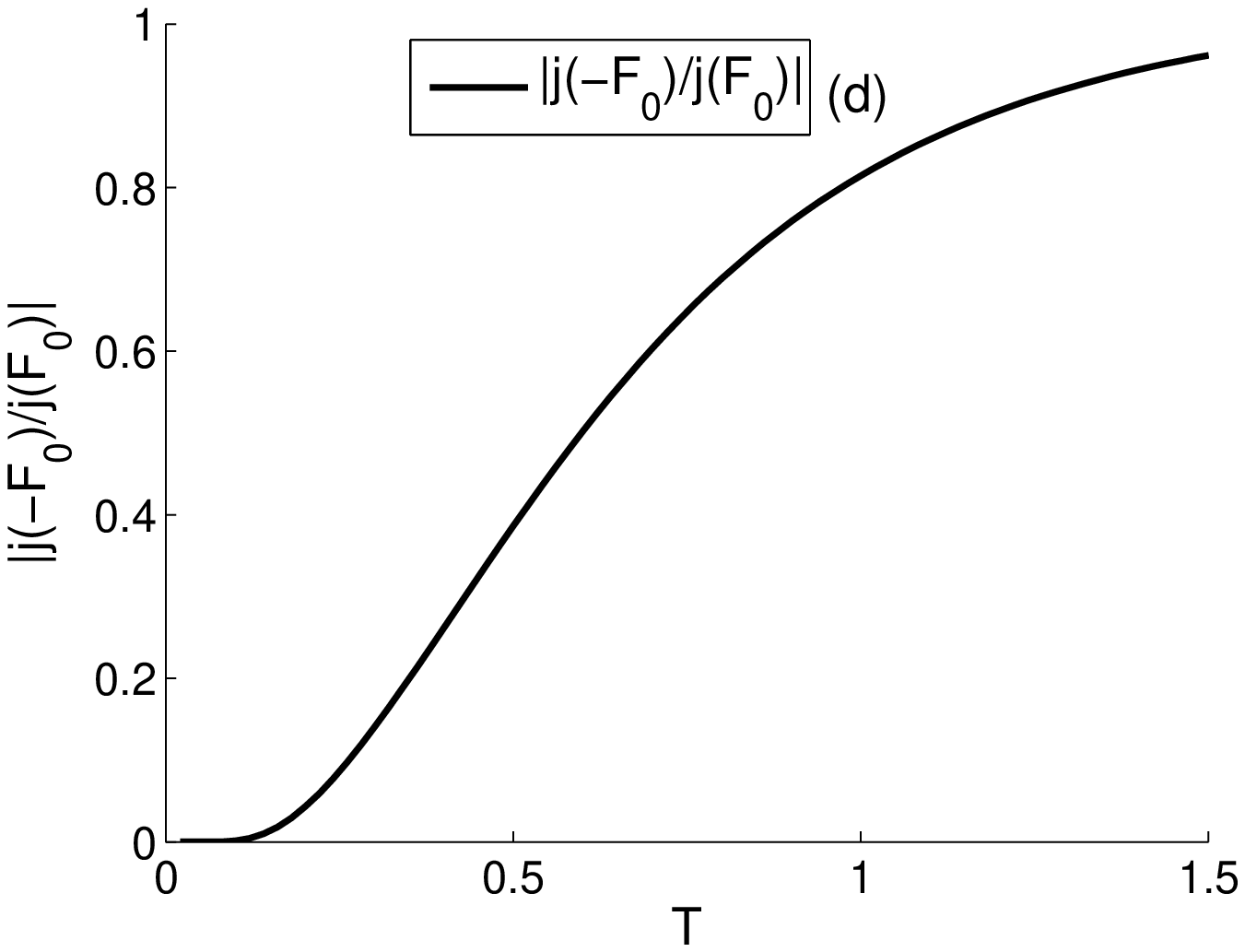}
  \caption{Plot of currents $j(F_{0}), j(-F_{0})$, and
  $|\frac{j(-F_{0})}{j(F_{0})}|$ for both the present system and the conventional ratchet. The parameters is the same as in Fig. 5.
  (a) $j(F_{0})$ and $j(-F_{0})$ vs $T$ in present system;
  (b)$|\frac{j(-F_{0})}{j(F_{0})}|$ vs $T$ in present system;
  (c)$j(F_{0})$ and $j(-F_{0})$ vs $T$ in the conventional energetic
  ratchet; (d)$|\frac{j(-F_{0})}{j(F_{0})}|$ vs $T$ in the conventional energetic ratchet.
  }\label{1}
\end{center}
\end{figure}

\indent Figure 7 shows current $j(F_{0}), j(-F_{0})$, and
  $|\frac{j(-F_{0})}{j(F_{0})}|$ as a function of temperature for both the present system and the conventional ratchet. The
  parameters is the same as in Fig. 5.  When a constant biased force acts
  on the motor [Fig. 7(a)], its current decreases monotonically with
  temperature which is contrary to the case of energetic barriers [Fig. 7(c)] \cite{15,17}.
In the presence of energetic barriers [Fig. 7(c)], the temperature
facilitates the activation and thus tends to increase the particle
current. However, in the presence of entropic barriers [Fig. 7 (a)],
the temperature dictates the strength of the entropic barriers and
thus an increasing temperature leads to reduction of the current.
 Equation (15) shows that the efficiency $\eta$ depends on
  $|\frac{j(-F_{0})}{j(F_{0})}|$. Figure 7 (b) shows that in the presence of entropic barriers the
  ratio $|\frac{j(-F_{0})}{j(F_{0})}|$ displays a clear minimum at
  the same value of temperature which corresponds to a maximum of
  $\eta$ in Fig. 5(a). However, in conventional ratchet the ratio
  $|\frac{j(-F_{0})}{j(F_{0})}|$ is a monotonically increasing
  function of temperature [Fig. 7 (d)] and thus the efficiency is a
  monotonically decreasing function of temperature [Fig. 5(b)]. So
  the thermal noise cannot facilitate energy transformation in
  this case.

 \section{Concluding Remarks}

\indent In this paper, we study the energetics of a Brownian
particle moving along the axis of a three-dimensional asymmetric
periodic tube in the presence of a symmetric unbiased force and a
load. The movement of the Brownian particle can be described by
Fick-Jacobs equation which is derived from 3D (or 2D) Smoluchowski
equation after elimination of $y$ and $z$ coordinates. Reduction of
coordinates may induce the appearance of entropic barriers and an
effective diffusion coefficient. In order to compare with the
conventional energetic ratchet on the quasistatic limit, the current
and the efficiency are also investigated in the presence of energy
barriers. The energetics in the presence of entropic barriers is
different from that occurring through energy barriers. It is found
that even on the quasistatic limit the efficiency can be a peaked
function of temperature, which proves that thermal noise can
facilitate energy transformation, contrary to the conventional
energetic ratchet \cite{15}. Two factors are essential for the
energy transformation activated by thermal noise: first one is the
noise-induced flow, and the second one is the finite dissipation in
the absence of thermal noise. For the case of entropic barriers,
even when the current goes to zero at $T=T_{c}$, local current still
remains finite. The input energy is a monotonically decreasing
function of temperature. Therefore at the limit $T\rightarrow 0$,
the input energy $E_{in}$ tends to its maximum, where all input
energy dissipates. The useful work takes its maximum at a finite
temperature. Thus, the efficiency can be a peaked function of
temperature. It is to be noted that the condition of maximum current
does not correspond to maximum efficiency, which is similar to the
case of energy barriers \cite{4,17}.\\

\begin{center}
    \textbf{{ACKNOWLEDGMENTS}}
\end{center}
 \indent The work was supported by the National
Natural Science Foundation of China under Grant No. 30600122 and
GuangDong Provincial Natural Science Foundation under Grant No.
06025073.

\end{document}